\documentclass[prl,showpacs,twocolumn]{revtex4-1}

\usepackage{amsmath}
\usepackage{graphicx}
\usepackage{bm}

\newcommand{\spin}{\sigma}

\begin{document}

\title{Energy and contact of the one-dimensional Fermi polaron at zero and finite temperature}
\author{E. V. H. Doggen and J. J. Kinnunen}
\affiliation{Department of Applied Physics, Aalto University School of Science, Finland}
\begin{abstract}
We use the T-matrix approach for studying highly polarized homogeneous Fermi gases in one dimension with repulsive or attractive contact interactions. Using this approach, we compute ground state energies and values for the contact parameter that show excellent agreement with exact and other numerical methods at zero temperature, even in the strongly interacting regime. Furthermore, we derive an exact expression for the value of the contact parameter in one dimension at zero temperature. The model is then extended and used for studying the temperature dependence of ground state energies and the contact parameter.
\end{abstract}

\maketitle

Strongly interacting Fermi gases are of interest in many fields of physics, such as atomic and nuclear physics, as well as the study of neutron stars and quantum chromodynamics. 
Furthermore, a better understanding of Fermi gases is essential for improving our description of elusive condensed matter phenomena, such as high-temperature superconductivity and (anti)ferromagnetism. 
The Fermi polaron is part of a class of phenomena known as impurity problems, which occur in a wide range of physical disciplines.

Recently it was shown in a series of papers by Tan \cite{Tan2008a,Tan2008b}, building on important early contributions provided by Olshanii and Dunjko \cite{Olshanii2003a}, that an interacting Fermi gas with short-range interactions exhibits a universal decay of the momentum distribution $n_k \sim C/k^4$ as $k \rightarrow \infty$.
The quantity $C$ is called the \textit{contact} and is intimately connected to the thermodynamics of the many-body system. 
In general it depends on physical properties such as temperature, pressure and scattering length. 
However, the universal decay holds for any value of these quantities as long as the interactions are short-range (for a review on Tan's contact parameter, see \cite{Zwerger2012a}). 
This behavior persists even in one dimension \cite{Barth2011a}. 
The contact has also been probed experimentally in Fermi gases \cite{Stewart2010a,Kuhnle2010a} and a Bose-Einstein condensate \cite{Wild2012a}. 
Since the contact of a homogeneous Fermi gas can be measured \cite{Jin2012a}, this provides an excellent test of theories. In particular, our work is relevant for a 1D quantum wire geometry \cite{Bloch2008a}.

The ground state energy of a polaron, a particle with generalized spin down interacting with a bath of spin up-particles (in experimental practice, the two different ``spin'' states are generally different hyperfine states), is predicted surprisingly well using a simple variational Ansatz due to Chevy \cite{Chevy2006a}. 
This work was later clarified and extended to other dimensions \cite{Combescot2007a,Giraud2009a,Massignan2011a} and applied to a lattice geometry \cite{Leskinen2010b}. 
Another simple approach gives reasonable results in all dimensions \cite{Klawunn2011a}. 
In 1D it is well known that the usual Fermi liquid picture breaks down due to the Peierls instability, and the quasiparticle picture fails due to the importance of collective excitations \cite{Voit1995a}. 
However, in the case of polarons the disturbance of the ideal Fermi distribution of the atoms of the majority component is negligible in the thermodynamic limit and we expect the diagrammatic approach to be applicable.
Such an approximative approach may seem superfluous given the exact results available from the Bethe Ansatz (BA) solution, but the T-matrix approach is simpler, more physically intuitive, easily extended to finite temperature and can be generalized to higher dimensions, where no BA solution is available.

We apply the non-self-consistent (NSC) T-matrix approach to the problem of a Fermi polaron in 1D, at zero as well as finite temperatures and for repulsive as well as attractive interactions. 
In doing so, we compute the ground state energy of the polaron and the contact density. 
Few exact solutions are available for fermionic many-body systems; we derive an exact expression for the contact density based on the BA solution, valid at zero temperature.
We compare our results to a more elaborate semi-self-consistent scheme based on the Brueckner-Goldstone theory \cite{Brueckner1955a,Goldstone1957a}, which has been widely used in the context of nuclear matter \cite{FetterAndWalecka}, and (perhaps counterintuitively) find that the self-consistent approach is less accurate than the NSC approach.

We consider a system of free fermions in 1D with a ``spin down'' particle interacting through a $\delta$-function potential with a sea of ``spin up'' particles. 
To wit, we use an interparticle potential (we set $\hbar = 1$) $V(x) = -\frac{2}{ma} \delta(x)$, where $m$ is the mass of a particle and $a$ is the 1D scattering length. 
The external potential is set to zero.
The system is thus described by the Hamiltonian:
\begin{align}
 & \mathcal{H} = \sum_\spin \int dx \psi^\dagger_\spin(x) \Big(-\frac{1}{2m}\frac{d^2}{dx^2} - \mu_\spin \Big) \psi_\spin(x) \nonumber \\
  & + \int dx \int dx' \psi^\dagger_\uparrow(x) \psi^\dagger_\downarrow(x') V(x-x') \psi_\downarrow(x') \psi_\uparrow(x),
\end{align}
where $\spin$ is the spin index, $\mu_\spin$ is the chemical potential and $\psi_\spin^{(\dagger)}(x)$ destroys (creates) a particle with spin $\spin$ at position $x$. 

The off-shell many-body T-matrix describes all possible (relevant) scattering events involving at most a single particle-hole-pair. 
It can be obtained numerically from the two-body T-matrix, which is easy to calculate for a $\delta$-function potential. 
In momentum space, this potential is given by a constant and the two-body T-matrix $\Gamma_0(S)$ can be obtained either from the Lippman-Schwinger equation \cite{Morgan2002a} or from a scattering calculation using the scattering amplitude $f(k) = \frac{-1}{1+iak}$ ($k$ is momentum) \cite{Olshanii1998a}, with identical results. 
The many-body T-matrix $\Gamma(S)$, which in the case of a $\delta$-potential depends only on the center-of-mass $1+1$-momenta (momentum and frequency), is then used according to the Galitskii integral approach in the ladder approximation to obtain the self-energy:
\begin{equation}
  \Sigma(P) = -i \int \frac{dK}{(2\pi)^2} G_0(K) \Gamma(S),  \label{selfenergy}
\end{equation}
where $K = \{k,\omega\}$ is the generalized $1+1$-momentum, $G_0(K)$ is the noninteracting Green's function for the majority component, and $S = \frac{1}{2}(P+K)$ are the center-of-mass variables. 
The ground state energy is obtained by starting with a guess for the polaron energy $E_p$ and then iterating equation (\ref{selfenergy}) using $E_p = \epsilon_p + \Sigma(p,E_p)$ (where $\epsilon_p = \frac{p^2}{2m}$ is the kinetic energy) to obtain a stationary solution.
The Green's functions entering the T-matrix in the NSC theory are undressed, hence the iterative loop in eq.\ (\ref{selfenergy}) does not alter the T-matrix.
As we shall see, improving upon the NSC theory is not easy and it is a better approximation than what may be expected at first glance.
To generalize this approach to finite temperatures, consider the noninteracting Green's function:
 \begin{equation}
  G_0(k,\omega) = \frac{\theta(k - k_\text{F})}{\omega - \epsilon_{k} + i\eta} + \frac{\theta(k_\text{F} - k)}{\omega - \epsilon_{k} - i\eta},
 \end{equation}
where $\theta$ is the step function as usual, $\eta \rightarrow 0^+$ and $k_\text{F}$ is the Fermi momentum of the majority component. 
In the thermodynamic limit, we can simply replace the step function by the Fermi-Dirac distribution at finite temperature.

\begin{figure}[!htb]
 \centering
 \includegraphics[width=\columnwidth]{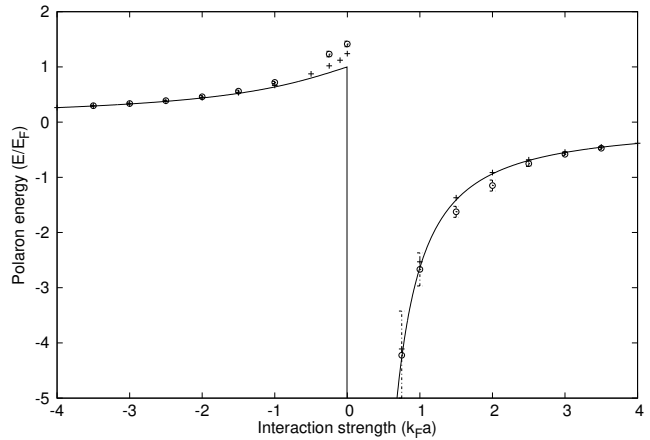}
 \caption{Energy of the ground state of the 1D Fermi polaron as a function of the interaction strength $k_\text{F}a$. Solid line shows the exact (BA) result, crosses (NSC) and circles (Brueckner-Goldstone theory) are numerical results from iterating eq.\ (\ref{selfenergy}) at various values of $k_\text{F} a$. Error estimates for the Brueckner theory results are related to poor convergence.} \label{polenergy}
\end{figure}

We compute the polaron ground state energy $E_{p = 0}$ in 1D by numerically evaluating $\Sigma$ at zero momentum over a large range of interaction strengths centered around infinitely strong interactions, i.e. $k_\text{F}a = 0$ (see Figure \ref{polenergy}).
Our results show good agreement with exact results obtained from the BA \cite{McGuire1965a,Mcguire1966a} as well as numerical results in the infinitely repulsive limit $k_\text{F}a \rightarrow 0^-$ obtained from a variational Ansatz \cite{Giraud2009a}. 
In fact, in this limit the NSC theory yields $E_{\text{NSC}}/E_\text{F} \approx 1.24$ (for $k_\text{F}a = -10^{-6}$), which is closer to the exact result of $E_{\text{ex}}/E_\text{F} = 1$ than the value reported in \cite{Giraud2009a} of $E_\text{var}/E_\text{F} \approx 1.39$. 
So while it was argued that in 3D the first order variational Ansatz of Chevy and the T-matrix approach are the same at zero temperature \cite{Combescot2007a}, this does not appear to be the case in 1D.
We argue that this feature is related to the fact that in 1D there is no ultraviolent divergence or a renormalization requirement, so that scattering of holes within the Fermi sphere is still relevant in 1D.
In the attractive regime, our numerics converge for $k_\text{F}a \gtrsim 0.5$.
At smaller (positive) values of the interaction strength our approach breaks down, possibly due to numerical instabilities.

We have also calculated the ground state energies using the Brueckner-Goldstone theory \cite{Brueckner1955a,Goldstone1957a}, which adds a self-consistency requirement to the T-matrix. 
The result is significantly worse than the NSC theory (see Figure \ref{polenergy}). 
This lends credence to the idea put forth in \cite{Combescot2008a}, where the authors argued that the self-consistent treatment performs poorly in highly spin-imbalanced systems because of the near-perfect destructive interference between diagrams that are not included in the NSC treatment. 
Extending to a self-consistent treatment only includes \textit{some} of these diagrams, and the destructive interference is lost.

\begin{figure}[!htb]
 \centering
 \includegraphics[width=\columnwidth]{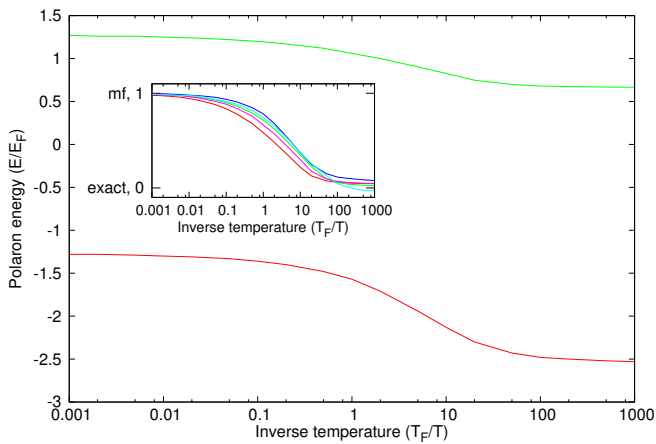}
 \caption{(Color online) Energy of the ground state of the Fermi polaron as a function of inverse temperature $T_\text{F}/T$ ($T_\text{F}$ is the Fermi temperature). Top (green) line shows the repulsive polaron at $k_\text{F}a = -1$, bottom (red) line shows the attractive polaron at $k_\text{F}a = 1$. Inset: normalized energy $E' = \frac{E - E_\text{ex}}{E_\text{mf} - E_\text{ex}}$ for various values of $k_\text{F}a = -0.5,-1,-2,-10,1$, where ``mf'' is the mean-field prediction and ``exact'' is the ground state polaron energy at $T = 0$ for the considered scattering length. The bottom (red) line is for $k_\text{F}a = -0.5$.} \label{polenergytemp}
\end{figure}

At finite temperature (see Figure \ref{polenergytemp}), we find that in the high-temperature limit, the mean-field limit $E_{\text{mf}}/E_\text{F} = -4/\pi k_\text{F}a$ is recovered, as expected. 
Furthermore, as $|k_\text{F}a|$ is reduced, it appears that higher temperatures are needed to reach the crossover from the BA-regime to the mean-field regime.

The Tan's contact is a central quantity connecting various thermodynamic quantities (such as energy, pressure and entropy) in atomic gases with short range interactions. 
It can be obtained from the asymptotic momentum distribution $n_k$ as $C = \lim_{k\rightarrow \infty} k^4 n_k$ and is the same for both the minority and the majority component.
Since the effect of the impurity on the momentum distribution of the majority component is vanishingly small, we can use the Fermi-Dirac distribution for the majority component, even though this does not have a $1/k^4$ tail. 
The many-body T-matrix can be used for calculating the contact from a perturbative approach as \cite{Sartor1980a,Kinnunen2012a}:
\begin{equation}
  C_\text{pert} = \Big(\frac{m}{L}\Big)^2 \sum_k \Big|\Gamma \Big(\frac{k}{2},\frac{\epsilon_{\downarrow,0} + \epsilon_{\uparrow,k}}{2} \Big)\Big|^2 n_{\uparrow,k} \label{perturbation},
\end{equation}
where $\epsilon_{\downarrow,0}$ is the polaron ground state energy and $\epsilon_{\uparrow,k} = \frac{k^2}{2m}$.
Alternatively, we can also compute the contact both numerically and exactly by evaluating the derivative of the ground state energy with respect to the scattering length, i.e. using the 1D Tan adiabatic theorem \cite{Barth2011a}:
\begin{equation}
 \frac{dE}{da} = \frac{C}{4m_r}, \label{1dtanadiabatic}
\end{equation}
where $m_r = m/2$ is the reduced mass.
Based on this theorem, we can use the exact BA ground state energies to calculate the exact dimensionless contact density $\mathcal{C}_\text{ex} = C/(Lk_\text{F}^4)$, where $L = \pi/k_{\text{F},\downarrow}$ is the length of the one-dimensional gas and $k_{\text{F},\downarrow}$ is the Fermi momentum of the minority component. The result is:
\begin{equation}
 \mathcal{C}_\text{ex} = \frac{4}{\pi^2 (k_\text{F}a)^3} \Big(k_\text{F} a + \arctan(\frac{1}{k_\text{F} a}) + \frac{\pi}{2} \Big). \label{exactcontact}
\end{equation}
In the infinitely repulsive limit, $k_\text{F} a \rightarrow 0^-$, this approaches the finite value of $4/3\pi^2$, which compares to the value $0.749...$ for the spin-balanced case \cite{Barth2011a}, and is exactly the same as the value obtained for a Tonks-Girardeau (TG) gas of infinitely repulsive one-dimensional bosons \cite{Gangardt2003a}.
Thus, an infinitely strongly repulsive impurity has the same contact when interacting with a ``fermionized'' sea of bosons as with an ideal Fermi sea. 
This is a somewhat surprising result considering that the momentum distributions of the TG Bose gas and the ideal Fermi gas are quite different, but one should observe that the impurity and the ideal Fermi gas actually constitute a TG Fermi gas \cite{Guan2009a}.
\begin{figure}[!htb]
 \centering
 \includegraphics[width = \columnwidth]{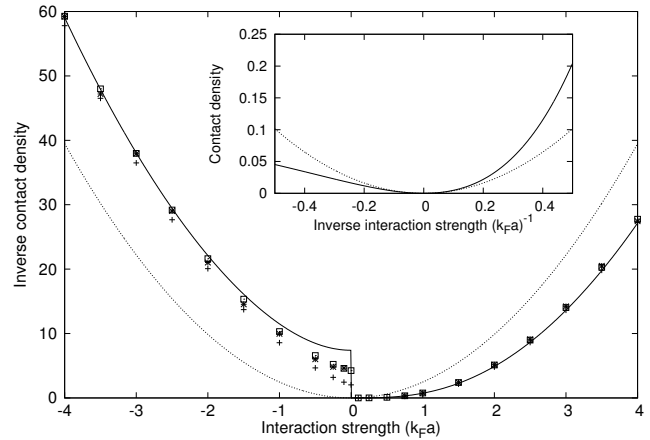}
 \caption{Inverse contact density $1/\mathcal{C}$ as a function of interaction strength at zero temperature. Solid line: exact result (\ref{exactcontact}). Dotted line: mean-field prediction. Crosses are numerical results obtained using eq.\ (\ref{perturbation}), squares are obtained from a modified version of eq.\ (\ref{perturbation}) (see main text) and stars are obtained from an implementation of the 1D Tan adiabatic theorem (\ref{1dtanadiabatic}) using numerically obtained values for the polaron energy. Inset: exact and mean-field values of the contact density around the weakly interacting limit $(k_\text{F}a)^{-1} = 0$.}
 \label{contact}
\end{figure}

Our results (see Figure \ref{contact}; $1/\mathcal{C}$ is shown to be able to plot the repulsive and attractive side in the same picture) show good agreement with the exact relation (\ref{exactcontact}). 
Eq.\ (\ref{perturbation}) works well in the weakly to moderately interacting regime, although the agreement deteriorates at stronger interactions.
Using the numerical derivative and eq.\ (\ref{1dtanadiabatic}) gives better results.
At $k_\text{F}a = -1$ we obtain for the exact value of the contact $\mathcal{C}_\text{ex} = (4-\pi)/\pi^2 = 0.0870...$, while using numerical values for the polaron energy and eq.\ (\ref{1dtanadiabatic}) gives $\mathcal{C}_\text{NSC} \approx 0.100$ and eq.\ (\ref{perturbation}) gives $\mathcal{C}_\text{pert} \approx 0.117$.
At $k_\text{F}a = 1$ we obtain $\mathcal{C}_\text{ex} = (4+3\pi)/\pi^2 = 1.36...$, $ \mathcal{C}_\text{NSC} \approx 1.28$ and $\mathcal{C}_\text{pert} \approx 1.57$.
No phase transition is found at any $k_\text{F}a$; the jump in $\mathcal{C}$ at $k_\text{F}a = 0$ is not a phase transition since the infinitely attractive and infinitely repulsive regime are not adiabatically connected.
This is a peculiarity of the 1D system; in the 3D case \cite{Punk2009a} there is a first order transition from a polaronic to a molecular state.
The perturbative theory (\ref{perturbation}) assumes the polaron to be in the zero momentum state. 
Clearly, at stronger interactions, the polaron has an increasingly high probability to be at higher momenta, and eq.\ (\ref{perturbation}) becomes increasingly inaccurate. Correcting for this improves the results (squares in Figure \ref{contact}) and makes them approximately in line with numerical results based on eq.\ (\ref{1dtanadiabatic}) (stars in Figure \ref{contact}); details of this method are shown in the supplementary material.
At $k_\text{F}a = -1$ this modified theory gives $\mathcal{C}_\text{mod} \approx 0.0970$, while at $k_\text{F}a = 1$ we obtain $\mathcal{C}_\text{mod} \approx 1.29$.
The difference between $\mathcal{C}_\text{mod}$ (squares) and $\mathcal{C}_\text{NSC}$ (stars) is due to an additional error from discretizing the derivative in eq.\ (\ref{1dtanadiabatic}).
The remaining error between $\mathcal{C}_\text{mod}$ and $\mathcal{C}_\text{ex}$ is due to the NSC T-matrix approximation.

\begin{figure}[!htb]
 \centering
 \includegraphics[width = \columnwidth]{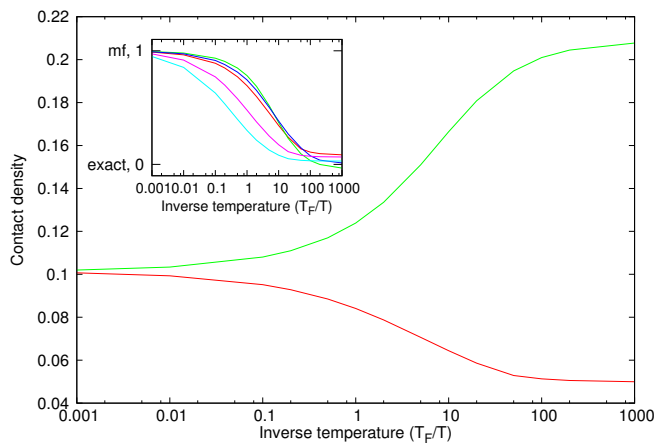}
 \caption{(Color online) Contact density $\mathcal{C}$ as a function of inverse temperature $T_\text{F}/T$. Top (green) line shows the attractive polaron at $k_\text{F}a = 2$. Bottom (red) line shows the repulsive polaron at $k_\text{F}a = -2$. The limiting value at high temperature, where both lines meet asymptotically, is the same as the value for the contact indicated by the dotted line in Figure \ref{contact}. Inset: normalized contact density $\mathcal{C}' = \frac{\mathcal{C} - \mathcal{C}_\text{ex}}{\mathcal{C}_\text{mf} - \mathcal{C}_\text{ex}}$ for various values of $k_\text{F}a = -0.25,-0.5,-2,2,-10$, where ``mf'' is the mean-field prediction and ``exact'' is the contact density at $T = 0$ for the considered scattering length. The bottom (cyan) line is for $k_\text{F}a = -0.25$. All values were calculated using perturbation theory (\ref{perturbation}).}
 \label{contacttemp}
\end{figure}

At finite temperature (see Figure \ref{contacttemp}), the contact reaches the mean-field regime at high temperatures. 
Interestingly, the contact reaches the same value regardless of the sign of $k_\text{F}a$ in this limit. 
Therefore, physically, it does not matter whether one considers attractive or repulsive interactions in the high-temperature limit, as thermal excitations dominate.
This feature can be understood upon inspection of eq.\ (\ref{perturbation}). 
As $T \rightarrow \infty$, the density distribution will become spread out over higher momenta, and the value of the T-matrix at higher momenta, which corresponds to the mean-field behavior of the system, will become increasingly important.
Another interesting feature is that for $k_\text{F}a < 0$, the contact density increases as the temperature increases.
This is a 1D feature that does not appear in the 3D spin-balanced case \cite{Jin2012a}.
Once again, we find that as interactions increase, it takes higher temperatures to reach the mean-field regime.
Physically, this means that while the mean-field prediction is a good approximation at high temperature for all values of $k_\text{F}a$, it takes increasingly high temperatures to reach this regime as $|k_\text{F}a|$ is reduced.

In conclusion, we have computed numerical approximations to the 1D Fermi polaron ground state energies and contact densities over a large range of interaction strengths. 
In addition, we have derived an exact expression for the contact density for the Fermi polaron in 1D at zero temperature. 
Our numerical approximations show good agreement with exact results. 
Finally, we have computed numerical results for finite temperatures and show a crossover to the mean-field regime at high temperatures.

\bibliographystyle{apsrev4-1}
\bibliography{ref}

\end{document}


SUPPLEMENTARY MATERIAL TO THE LETTER: ``Energy, contact and universal thermodynamics of the one-dimensional Fermi polaron at finite temperature''
\section{Contact from perturbation theory}

In Ref. \cite{Kinnunen2012a} one of the authors gave a perturbative result for calculating the probability of finding the polaron in momentum state $p$ (i.e. the polaron momentum distribution $\nu_p$) using the scattering T-matrix $\Gamma$. The one-dimensional analog of this result is
\begin{equation}
  \nu_p = \sum_q \left|\Gamma\left(\frac{E + \epsilon_q}{2},\frac{q}{2}\right)\right|^2 \left(\frac{1}{E+\epsilon_{q}-\epsilon_{q-p} -\epsilon_{p}}\right)^2 n_q (1-n_{q-p}),
\label{eq:pertmom}
\end{equation}
where $E$ is the polaron energy and $n_q$ is the majority component occupation number for momentum state $q$ (we will here assume zero-temperature distributions and set $\hbar = 1$ for the sake of simplicity).
The contact can be obtained from the high momentum limit of the momentum distribution as
\begin{equation}
C_\mathrm{pert} = \lim_{p\rightarrow \infty} p^4 \nu_p = \frac{m^2}{L^2} \sum_q \left|\Gamma\left(\frac{E + \epsilon_q}{2},\frac{q}{2}\right)\right|^2 n_q,
\end{equation}
where in the non-self-consistent T-matrix theory
\begin{equation}
\Gamma(q_0,q) = \frac{g}{1-g\sum_{|k'|>\kf} \frac{1}{2q_0-\epsilon_{k'} -\epsilon_{2q-k'}}}.
\end{equation}
Here $g = -2/ma$ (this is the prefactor in front of the delta function in the interparticle potential) and the sign convention implies that $a$ is negative for repulsive interactions. As we shall soon see, this form for the contact is, up to a normalization, identical to the contact obtained from the first-order variational Chevy ansatz and in agreement with the Tan's adiabatic theorem.

\section{Contact from Chevy Ansatz}
The first-order variational Chevy Ansatz \cite{Chevy2006a}, in which only a single particle-hole excitation is considered, is
\begin{equation}
|\psi\rangle = \phi_0 b_0^\dagger |0\rangle + \sum_{kq} \phi_{kq} b_{q-k}^\dagger c_k^\dagger c_q |0\rangle,
\end{equation}
where the vacuum $|0\rangle$ corresponds to the ideal Fermi sea of majority atoms. The first term describes a zero-momentum polaron and the second term has the polaron (minority atom) in the momentum state $q-k$. The coefficients $\phi_{kq}$ will give the polaron momentum distribution as
\begin{equation}
  \nu_p = \langle \psi|b_p^\dagger b_p|\psi\rangle = \sum_q \left|\phi_{q-p,q}\right|^2 n_{q} \left( 1-n_{q-p} \right).
\label{eq:polmom}
\end{equation}
Solving the polaron momentum distribution is quite straightforward from the Chevy ansatz if one neglects the hole-hole scattering term, leading to equations \cite{Giraud2009a}
\begin{equation}
 g\sum_{|k|>\kf,|q|<\kf} \phi_{kq} = E \phi_0-g\phi_0 n_\uparrow
\label{eq:ansatz1}
\end{equation}
and
\begin{equation}
 g \phi_0 + g\sum_{|k'|>\kf} \phi_{k'q} = \left( E+\epsilon_q-\epsilon_k -\epsilon_{q-k} \right) \phi_{kq},
\label{eq:ansatz2}
\end{equation}
where $n_\uparrow$ is the density of the majority component.
From Eq.~\eqref{eq:ansatz2} we obtain
\begin{equation}
\phi_{kq} = \frac{g \phi_0 + g\sum_{|k'|>\kf} \phi_{k'q}}{E+\epsilon_q-\epsilon_k -\epsilon_{q-k}}.
\label{eq:phi}
\end{equation}
Denoting the nominator in the right as $\chi(q) = g \phi_0 + g\sum_{|k'|>\kf} \phi_{k'q}$ yields
\begin{equation}
  \chi(q) = g\phi_0 + g\sum_{|k'|>\kf} \frac{\chi(q)}{E+\epsilon_q-\epsilon_{k'} -\epsilon_{q-{k'}}},
\end{equation}
which can be solved as
\begin{equation}
  \chi(q) = \frac{g\phi_0}{1- g\sum_{|k'|>\kf} \frac{1}{E+\epsilon_q-\epsilon_{k'} -\epsilon_{q-{k'}}}} = \phi_0 \Gamma\left(\frac{E+\epsilon_q}{2},\frac{q}{2}\right).
\label{eq:chi}
\end{equation}
The polaron momentum distribution can now be expressed by combining Eqs.~\eqref{eq:polmom},~\eqref{eq:phi}, and~\eqref{eq:chi}:
\begin{equation}
  \nu_p = \sum_q \left|\phi_0 \Gamma\left(\frac{E+\epsilon_q}{2},\frac{q}{2}\right) \frac{1}{E+\epsilon_q-\epsilon_{q-p} -\epsilon_{p}} \right|^2 n_q \left( 1-n_{q-p} \right).
\label{eq:finalmom}
\end{equation}
The expression on the right is the perturbative result in Eq.~\eqref{eq:pertmom} except for the factor $|\phi_0|^2$. Indeed, the perturbative result is built on the assumption that the polaron is (with unit probability) in the zero-momentum state but the perturbative correction will yield an unnormalized state. Renormalizing the perturbed state would yield the correct prefactor and the exactly same result as in Eq.~\eqref{eq:finalmom}. 
Notice that by substituting Eq.~\eqref{eq:chi} into Eq.~\eqref{eq:ansatz1} would yield the polaron energy
\begin{equation}
 E = \sum_q \Gamma(\frac{E+\epsilon_q}{2},\frac{q}{2}) n_q
\end{equation}
which is precisely the polaron energy obtained from the non-self-consistent T-matrix theory.

\section{Contact from energy}
The polaron energy, as computed from the Chevy ansatz, is
\begin{equation}
  E(a) = \sum_{q < \kf} \frac{g}{1 - g\sum_{k > \kf} \frac{1}{E(a) + \epsilon_q - \epsilon_k - \epsilon_{q-k}}} = \sum_{q < \kf} \frac{g}{1 - g\chi(E(a),q)}
\end{equation}
where $g = -2/ma$ and energy $E(a)$ will depend on scattering length $a$.
According to the one-dimensional analog of the Tan adiabatic theorem as derived by Barth and Zwerger \cite{Barth2011a}
\begin{equation}
  C = 2m \frac{\partial E(a)}{\partial a} = \frac{\partial}{\partial a} \left[ \sum_{q < \kf} \frac{g}{1 - g\chi(E(a),q)} \right].
\end{equation}
Calculating the derivative yields
\begin{equation}
  \frac{\partial E}{\partial a} = -\frac{1}{a} E(a) -  \sum_{q < \kf} \frac{g}{\left[1 - g\chi(E(a),q)\right]^2} \frac{\partial}{\partial a} \left[1-g\chi(E(a),q)\right],
\end{equation}
where we have used the property $\frac{\partial g}{\partial a} = -\frac{g}{a}$.
Now
\begin{equation}
\frac{\partial}{\partial a} \left[1-g\chi(E(a),q)\right] = \frac{g}{a} \chi(E(a),q) - g \frac{\partial \chi}{\partial a},
\end{equation}
which, upon substitution, yields
\begin{equation}
  \frac{\partial E}{\partial a} = -\frac{1}{a} E(a) -  \sum_{q < \kf} \frac{g}{\left[1 - g\chi(E(a),q)\right]^2} \left[\frac{g}{a} \chi(E(a),q) - g \frac{\partial \chi}{\partial a}\right].
\end{equation}
Combining the first and second terms on the right yields
\begin{equation}
  \frac{\partial E}{\partial a} = \frac{2}{m} \sum_{q < \kf} \left( \frac{g}{1-g\chi(E(a),q)}\right)^2 +  \sum_{q < \kf} \frac{g^2\frac{\partial \chi}{\partial a}}{\left[1 - g\chi(E(a),q)\right]^2}.
\end{equation}
The first term on the right is proportional to the perturbative result $C_\mathrm{pert}$ that we have been using in the main paper, the second term will yield the normalization as will be seen in a moment.
Evaluating the differential on the second term yields
\begin{equation}
\frac{\partial \chi(E(a),q)}{\partial a} =\frac{\partial}{\partial a} \sum_{k > \kf} \frac{1}{E(a) + \epsilon_q - \epsilon_k - \epsilon_{q-k}} = \sum_{k > \kf} \frac{\frac{-\partial E(a)}{\partial a}}{\left[E(a) + \epsilon_q - \epsilon_k - \epsilon_{q-k}\right]^2}.
\end{equation}
Thus we now obtain
\begin{equation}
  C = C_\mathrm{pert} -  2m \frac{\partial E(a)}{\partial a} \sum_{q < \kf} \frac{g^2}{\left[1 - g\chi(E(a),q)\right]^2}\sum_{k > \kf} \frac{1}{\left[E(a) + \epsilon_q - \epsilon_k - \epsilon_{q-k}\right]^2},
\end{equation}
rearranging the sums in the second term and substituting the value of the contact yields
\begin{equation}
  C = C_\mathrm{pert} - C \sum_{k > \kf} \sum_{q < \kf} \left[ \frac{g^2}{\left[1 - g\chi(E(a),q)\right]^2} \frac{1}{\left[E(a) + \epsilon_q - \epsilon_k - \epsilon_{q-k}\right]^2}\right],
\end{equation}
and further
\begin{equation}
  C = C_\mathrm{pert} - C \sum_{p \neq 0} \sum_q \left[ \frac{n_q (1-n_{q-p}) g^2}{\left[1 - g\chi(E(a),q)\right]^2} \frac{1}{\left[E(a) + \epsilon_q - \epsilon_{q-p} - \epsilon_p\right]^2}\right],
\end{equation}
where the term in the square brackets is the perturbative result for the probability for a majority atom scattering from momentum state $q$ to $q-p$. Of course, should that be the case, the polaron will then have scattered to momentum state $p$ and the summation over $q$ will then give the polaron occupation number $\nu_p$.
Thus we obtain
\begin{equation}
  C = C_\mathrm{pert} -  C \sum_{p \neq 0} \nu_p,
\end{equation}
which finally yields
\begin{equation}
  C = \frac{C_\mathrm{pert}}{1 + \sum_{p \neq 0} \nu_p}.
\end{equation}
That is, the perturbative result for the contact $C_\mathrm{pert}$ needs to be simply normalized by the polaron momentum distribution. In the weakly interacting limit, the probability for the polaron not being in the zero momentum state is very small, and the perturbative result without the normalization correction is accurate.